\documentclass[journal]{IEEEtran}

\usepackage{amsmath}
\usepackage{tabularx}
\usepackage{multirow}
\usepackage{graphicx}
\usepackage{epstopdf}
\usepackage{wrapfig}
\usepackage{graphicx}
\usepackage{xcolor}

\usepackage{setspace}
\usepackage{graphicx}
\usepackage{amsmath} 
\usepackage{mathtools}
\usepackage{amsfonts}
\usepackage{bbm}
\usepackage{braket}
\usepackage{caption}
\usepackage{graphicx}
\usepackage{float}
\usepackage{caption}
\usepackage{subcaption}
\usepackage{url}
\usepackage{color}
\usepackage[ruled]{algorithm2e}

\usepackage{amssymb}
\usepackage{cancel}
\usepackage{upgreek}
\usepackage{tabularx}
\usepackage{color}
\usepackage{units}
\usepackage{hhline}
\usepackage{textfit} 

\usepackage{slashed}
\graphicspath{ {images/} }

\usepackage{graphicx}
\usepackage{dcolumn}
\usepackage{bm}
\usepackage{hyperref}
\usepackage[mathlines]{lineno}

\usepackage{tikz}
\usetikzlibrary{positioning}
\usepackage{qcircuit}
\usepackage{braket}
\usepackage{indentfirst}
\usepackage{float}

\begin{document}

\title{Experimental Quantum Computing to Solve Network DC Power Flow Problem}

\author{Rozhin~Eskandarpour,~\IEEEmembership{Senior~Member,~IEEE},~Kumar~Ghosh,~\IEEEmembership{Member,~IEEE},~Amin~Khodaei,~\IEEEmembership{Senior~Member,~IEEE},~Aleksi~Paaso,~\IEEEmembership{Senior~Member,~IEEE}
        \vspace{-6mm}
\thanks{R. Eskandarpour and Kumar Ghosh and A. Khodaei are with the University of Denver, Denver, CO, USA. E. A. Paaso and S. Bahramirad are with Commonwealth Edison Company, Chicago, IL, USA.}
}

\maketitle

\begin{abstract}
Practical quantum computing applications to power grids are nonexistent at the moment. This paper investigates how a fundamental grid problem, namely DC power flow, can be solved using quantum computing. Power flow is the most widely used power system analysis technique, either as a stand-alone application or embedded in other applications; therefore, its fast and accurate solution is of utmost significance for grid operators. We base our studies on the Harrow-Hassidim-Lloyd (HHL) quantum algorithm, which has a proven theoretical speedup over classical algorithms in solving a system of linear equations. Practical studies on a quantum computer are conducted using the WSCC 9-bus system.     

\end{abstract}


\vspace{-3mm}
\section{Introduction}
\IEEEPARstart{T}HE POWER FLOW analysis is the keystone of electric utilities’ decision making in power grid operation, control, and planning. The solution of the power flow determines the amount of flow of electricity on various transmission and distribution lines, and voltage values on various nodes. Power flow is used in a wide range of applications from resource management, portfolio optimization, security analysis and grid upgrades, among other purposes. However, it is a daunting task to solve this problem accurately and efficiently due to nonlinearity induced by the laws of physics. 

The growing proliferation of clean renewable energy resources, the need for increased security and resilience against natural and human-made disasters, and growing expectations of electricity customers to high reliability and quality power have highlighted the importance of fast and accurate power flow studies more than ever. The traditional methods to solve the power flow problem involved reformulation, approximations or parallel processing to speed up its computation time. Unfortunately, applying these methods to emerging smart grids remains prohibitively ineffective due to a lack of mathematical convergence guarantee and unscalable analytics. 

In this paper, we propose to model, simulate, and experimentally solve the DC power flow problem on a quantum computer. A quantum computer can achieve enormous processing power by executing multiple functions simultaneously using possible permutations \cite{ber09}\cite{q0}; therefore, we expect a computational speedup over classical methods. This is discussed in detail through a sensitivity analysis of respective quantum and classical time complexities.

\vspace{-2mm}
\section{Quantum Algorithm For Solving DC Power Flow Problem}
\vspace{-1mm}

Solving a system of linear equations (SLE) is fundamental in many science and engineering problems. Quantum computers are proven to provide an exponential speedup for SLE \cite{qc1}\cite{qc2}.
The DC power flow analysis is an SLE that can be efficiently solved using a quantum computer. DC power flow can be illustrated as: given a matrix $B\in \mathbb{R}^{N\times N}$ and a vector $p\in \mathbb{R}^N$, find $\theta \in \mathbb{R}^N$ while satisfying $B\theta = p$. Here $B$ represents the admittance matrix, $p$ denotes nodal injections (positive if generation, and negative if load), and $\theta$ that are to be determined, represent the nodal voltage angles. This SLE is called s-sparse if $B$ has at most $s$ nonzero entries per row or column, which is the case for the DC power flow problem. Solving an s-sparse system of size $N$ with a classical computer requires $O\left(Nsk       \log\left(\frac{1}{\epsilon}\right)\right)$ running time using the conjugate gradient method $\cite{qc1}$. However, a quantum computer using the Harrow-Hassidim-Lloyd (HHL) quantum algorithm can achieve  $O\left(\log (N) \frac{s^2 k^2}{\epsilon} \right)$ running time \cite{hhl}. Here $k$ and $\epsilon$ denote the condition number of the system and the accuracy of the approximation, respectively. The HHL model of the DC power flow problem is achieved as discussed in the following. 

The DC power flow equations should be rescaled/normalized for quantum applications. We perform this by normalizing $p$ and $\theta$ and mapping to the respective quantum states $\ket{p}$ and $\ket{\theta}$. Matrix $B$ is already Hermitian (has a conjugate transpose similar to itself), so no change on $B$ is needed. The rescaled problem to be solved will be as follows: 
\begin{equation}
B  \ket{\theta} = \ket{p} \label{equationbthetap}
\end{equation}

We will use three types of quantum registers: The first register is used to store a binary representation of the eigenvalues of $B$ and is denoted by $\alpha$. The second register contains the vector solution indicated by $\beta$, where $N=2^\beta$. The third register is for ancilla qubits. All three registers are set to $\ket{0}$ at the beginning of each computation and restored to $\ket{0}$ at the end.

We initially load the data into $\ket{p}$, through the following transformation:
\begin{equation}
\ket{0}_\beta = \ket{p}_\beta
\end{equation} 
Once loaded, we will apply a Quantum Phase Estimation (QPE) to find the eigenvector of $B$. This is a controlled unitary with a change of basis that maps the eigenvalues onto the working memory. 
We will define QPE with $U=e^{iBt}$, where 
\begin{equation}
	e^{iBt} = \sum_{i} e^{i\lambda_j t} | u_j \rangle \langle u_j|,
\end{equation}
For the eigenvector $\ket{u}_\beta$, which has eigenvalue  $e^{i\lambda_jt}$, QPE will output $\ket{\widetilde{\lambda_j}}_\alpha\ket{u_i}_\beta$, where $\widetilde{\lambda}_j$ represents an $\alpha$-bit binary approximation to $2^\alpha \lambda_j t/2\pi$. If each $\lambda_j$ can be correctly represented with $\alpha$ bits, we will have
\begin{equation}
	QPE(e^{iB2\pi} \sum_j p_j| 0 \rangle_\alpha| u_j \rangle_\beta) = \sum_{j} p_j | \lambda_j \rangle \langle u_j|
\end{equation}     
The quantum state of the register stated in the eigenbasis of \textit{B} is:
\begin{equation}
	 \sum_{j} p_j \ket{\lambda_j}_\alpha \ket{u_j}_\alpha
\end{equation} 
where $\ket{\lambda_j}_\alpha$ is the $\alpha$-bit binary representation of $\lambda_j$. Now that the register is ready, we will add the ancilla qubit and apply a controlled rotation conditioned on  $\ket{\lambda_j}$:
\begin{equation}
	\sum_j p_j\ket{\lambda_j}_\alpha \ket{u_j}_\alpha \left(\sqrt{1-\frac{c^2}{\lambda_j^2}} \ket{0} + \frac{c}{\lambda_j} \ket{1} \right)
\end{equation}
where $c$ is a normalization constant. Now we can apply a reverse QPE to  uncompute the results in (6), which will result in
\begin{equation}
	\sum_j p_j \ket{0}_\alpha \ket{u_j}_\alpha \left(\sqrt{1-\frac{c^2}{\lambda_j^2}} \ket{0} + \frac{c}{\lambda_j} \ket{1} \right)
\end{equation}

Finally, we can measure the ancilla qubit. The register is in the post-measurement state when the outcome is 1:
\begin{equation}
	\sqrt{\frac{1}{\sum_j|p_j|^2/|\lambda_j|^2}} . \sum_j  p_j \ket{0}_\alpha \ket{u_j}_\alpha \label{ancilla}
\end{equation}

In the readout register we obtain a normalized quantum state $\ket{\psi} =\frac{\left|\theta \right\rangle }{\left|\left|\theta \right|\right|}$, where $\ket{\theta} = B^{-1} \ket{p}$ is the normalized solution of the DC power flow in (\ref{equationbthetap}). The normalization constant $\left|\left|\theta \right|\right|$ is obtained by measuring the ancilla qubit in (\ref{ancilla}).

\section{Numerical Studies}
\begin{figure}[htp]
    \centering
	\includegraphics[width= 3.3 in]{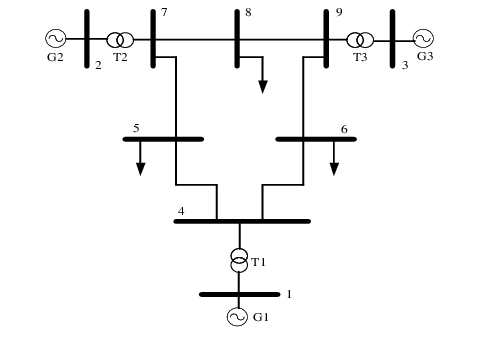}\\
	\caption{WSCC 9-bus test system.}
	\centering
	\label{fig:Fig3}
\end{figure}

The WSCC 9-bus system, shown in Fig. \ref{fig:Fig3}, is used for solving the DC power flow problem using the HHL algorithm. This system approximates the western US power grid, called the Western System Coordinating Council (WSCC), to an equivalent grid with nine nodes and three generators. The system data can be found in \cite{9bus}. The quantum circuit was designed with Qiskit \cite{Qasm}, an open-source software development kit for working with quantum computers at the circuits level, and simulated on the IBM Q Experience cloud quantum computing platform.  
\subsection{Implementation}
We attempt to select a proper value for $t$ to help improve the accuracy of the solution returned by HHL. This value would help rescale the eigenvalues during QPE. The fidelity of the solution should be near $1$  in an ideal case, however the noise added by the quantum hardware will result in the fidelity of less than $1$. Table \ref{table1} demonstrates the results of the studied DC power flow problem. The obtained fidelity is $0.87$. This high fidelity shows the promise in finding accurate solutions, despite the noise and the large size of the quantum circuit. 

\begin{table}[H]
\caption{Quantum Simulation Results} 
\centering 
\begin{tabular}{| c | c |} 
\hline 
& Results ~\\ [0.5ex] 
\hline 
Fidelity & \textbf{0.872159}  \\ \hline 
 Probability & 0.537650  \\ \hline 
 \shortstack{\newline Classical \\ Solution} & \shortstack{ [0.1157,  0.0948,  -0.0713,  -0.1316,\\ -0.0644,  0.0118, -0.0514,  0.0220]  }  \\  \hline 
 \shortstack{Quantum \\ Solution} & \shortstack{[0.1125,  0.0599, -0.0299, -0.0826,\\ -0.0458, 0.0047, -0.0858, -0.0040]  }  \\ 
\hline 
\end{tabular}
\label{table1} 
\end{table}

Table \ref{table2} shows the characteristics of the quantum circuit used to solve this problem. The depth is the maximum number of gates applied to a single qubit. The width is the required number of qubits. The number of CNOTs is also provided as this number, along with the width, provide a sense of whether running the circuit on current practical hardware is feasible.

\begin{table}[H]
\caption{ Characteristics of The Quantum Circuit} 
\centering 
\begin{tabular}{|c|c|} 
\hline 
& Results ~\\ [0.5ex] 
\hline 
Circuit Width & 9 \\ \hline
 Circuit Depth & 129  \\ \hline 
 CNOT Gates & 70  \\  
\hline 
\end{tabular}
\label{table2} 
\end{table}

\subsection{Time Complexity and Quantum Speedup}


The best known time complexity for the classical solution of the SLE is given by $O\left(N s k \log\left(\frac{1}{\epsilon}\right)\right)$, whereas the time complexity for the HHL algorithm is $O\left(\log (N) \frac{s^2 k^2}{\epsilon} \right)$. We calculate the parameters for the studied problem as follows:

\begin{itemize}
    \item[-] $N$ is the row/column dimension of the square matrix $B$, which is the number of system buses minus 1. This is true for larger systems as well. For the studied 9-bus system, $N$ is equal to 8.   
    \item[-] The HHL algorithm generally assumes that the singular values of $B$ lie between $1/k$ and $1$; for our system, and based on the eigenvalues of $B$, the condition number $k$ is calculated as $0.017$.
    \item[-] The sparsity, $s$, is the maximum number of nonzero elements in each row. For the studied system, $s$ is $4$, stemming from the fact that no node is connected to more than three other nodes (resulting in three non-diagonal elements and one diagonal element in $B$). The sparsity does not grow significantly with the system size, as the number of nodes connected to a single node is commonly limited in practical power grids. 
    \item[-] The approximation accuracy, $\epsilon$, is the additive error achieved in the output state, related to the desired precision. The HHL algorithm returns an approximate state $\ket{x^\prime}$ of the state $\ket{x}$ of the exact solution. This means $|\ket{x} - \ket{x^\prime} | \leq \epsilon$. Here we assume $\epsilon =\sqrt{2 (1 - \sqrt{\text{fidelity}})}$, where $\text{fidelity} = |\langle x^\prime | x \rangle|$. For the studied system, fidelity is obtained as $0.87$, which results in $\epsilon = 0.37$.  
\end{itemize}

Using the above-mentioned parameters and the theoretical time complexity for classical and quantum solutions, we can compare the performance of classical and quantum computers in solving the DC power flow problem. For the classical solution, the condition number and sparsity will be similar to the values discussed above. However, considering a much higher fidelity of $0.99$ to represent the maturity of the classical computing technology, we obtain $\epsilon = 0.1$.  

Now we have to make a base assumption of how long it would take for each technology to solve a $2\times2$ SLE, and then use the time complexities to rescale to the studied 9-bus test system. Plugging the numbers into time complexities, we observe that if the classical computer is $34$ times faster than the quantum computer in solving the $2\times2$ SLE, both technologies will solve the 9-bus system with the same speed. This faster speed in solving a base problem ties the simulation time to physical parameters, including the quantum hardware. Analogously, if we assume that the computation times for both methods are the same in solving the base case, the quantum computer will solve the DC power flow for the 9-bus system $34$ times faster than the classical computer. 

\begin{figure}[tp]
    \centering
	\includegraphics[width= 3.3 in]{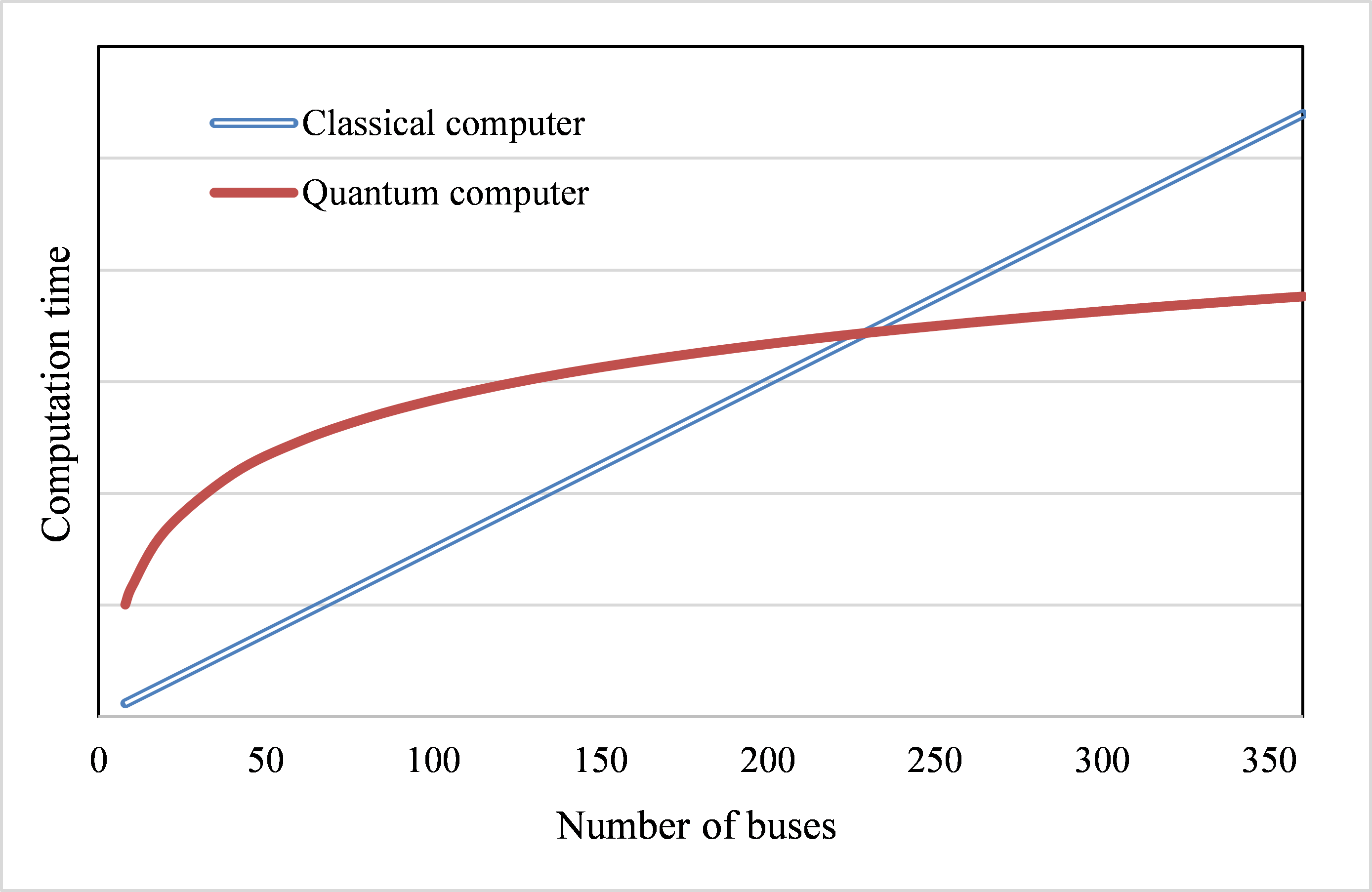}\\
	\caption{Comparison between classical and quantum computation time for a various number of buses.}
	\centering
	\label{fig:Fig4}
\end{figure}

We build upon this observation, considering a base computational speed difference of $34$ times, and re-study the time complexity of both technologies based on more conservative estimates. By increasing either the condition number or the sparsity, the quantum speed will be reduced more than its classical counterpart. Therefore, we consider $k=0.1$ and $s=6$ to ensure conservativity, and obtain the results shown in Fig. \ref{fig:Fig4} based on a various number of buses. 


The intersection of these the curves is at $N=230$, advocating that for $230$-bus and larger systems, the quantum computer will find the solution faster than a classical computer. Note this is based on conservative assumptions. This speedup will be more considerable for larger systems, as shown in the figure.  

\subsection{Discussions}
There are multiple points to consider regarding the obtained results: 
\begin{itemize}
    \item[-] The currently available quantum computers are noisy (fitting under the category of Noisy Intermediate-Scale Quantum (NISQ) computers), so we cannot yet expect an exact solution from this technology, even if tested on small-scale problems as discussed in this paper. 
    \item[-]     The discussed quantum speedup is theoretical. Given the limited size of existing quantum computers, experimental speedup comparisons are not possible at this point in time. 
    \item[-]     The proposed solution algorithm, i.e., HHL, may present scalability issues when the studied system becomes larger; however, the proposed extended studies were based on the assumption that this algorithm can be applied to larger test systems.  
\end{itemize}
Despite the more theoretical perspective, the studied system demonstrates an obvious superior quantum performance compared to classical solution in terms of computation time. Moreover, the mentioned challenges are expected to be overcome as the quantum computing technology progresses and matures. 

\vspace{-3mm}
\section{Conclusion}
This paper presented an experimental quantum computing solution to the widely-used DC power flow problem. The results showed that a quantum computer could provide higher processing power and solve this fundamental grid problem much faster than a classical computer. It was further shown that this conclusion holds true even for cases that quantum hardware adds delays to computations. We expect this successful experimental study to enable broader research and applications of quantum computing to power grids in the near future.

\ifCLASSOPTIONcaptionsoff
  \newpage
\fi

\bibliographystyle{IEEEtran}
%

\end{document}